\documentclass[showpacs, pra,twocolumn,preprintnumbers ,amsmath, amssymb, superscriptaddress, aps]{revtex4-2}
\usepackage{color}
\usepackage{amsmath,amssymb}
\usepackage{pifont}% http://ctan.org/pkg/pifont
\usepackage{amssymb}  % More symbols
\usepackage{bbold}
\usepackage{float}
\usepackage{subfloat}

\usepackage[caption=false]{subfig}
\usepackage{tikz}
\usepackage{makecell}
\usepackage{subfig}
\usepackage{pifont}   % Ding symbols
\usepackage{graphicx} % Include figure files
\graphicspath{{Figures/}}
\usepackage{dcolumn}  % Align table columns on decimal point
\usepackage{bm}       % bold math
\usepackage{multirow} % Table functions
\usepackage{placeins}% For making floats not move around everywhere
\usepackage[colorlinks]{hyperref}
\usepackage{mathtools}
\usepackage{appendix}

\begin{document}

    \title{Strain effect on Goos–Hänchen shifts and group delay time in gapped graphene
barrier}
    \date{\today}
    \author{Miloud Mekkaoui}
    %   \email{benlakhouy.n@ucd.ac.ma}
    \affiliation{Laboratory of Theoretical Physics, Faculty of Sciences, Choua\"ib Doukkali University, PO Box 20, 24000 El Jadida, Morocco}

    \author{Youssef Fattasse}
    %   \email{benlakhouy.n@ucd.ac.ma}
    \affiliation{Laboratory of Theoretical Physics, Faculty of Sciences, Choua\"ib Doukkali University, PO Box 20, 24000 El Jadida, Morocco}
 \author{Ahmed Jellal}
\email{a.jellal@ucd.ac.ma}
\affiliation{Laboratory of Theoretical Physics, Faculty of Sciences, Choua\"ib Doukkali University, PO Box 20, 24000 El Jadida, Morocco}
\affiliation{Canadian Quantum  Research Center,
    204-3002 32 Ave Vernon,  BC V1T 2L7,  Canada}

    \pacs{
    }
    \begin{abstract}
 We investigate the strain effect on the Goos–Hänchen (GH) shifts and
group delay time for transmitted Dirac fermions in gapped graphene
through a single barrier potential. The solutions of energy spectrum are used to compute the transmission probabilities together with the GH shifts and group delay time. Our  results show that the two last quantities
are strongly depending to weather the strain is applied along armchair or zigzag directions. In particular it found that   both of quantities can be enhanced with the applied  strain.

    \end{abstract}

    \maketitle
    \section{Introduction}
%Graphene \cite{Novoselov1, Geim}
%is an one-atom thick carbon material, which has
%shown very interesting electronic properties due to its massless
%Dirac electronic structure.
Since its successful isolation in  2004 \cite{Novoselov1, Geim},
 graphene has attracted a
considerable attention from both experimental and theoretical
investigations. This is because of its unique and outstanding
mechanical, electronic, optical, and thermal  properties
\cite{Castro}. On the other hand, there is a big progress in studying
quantum phenomena  in graphene systems among them we cite the
quantum version of the Goos-H\"anchen (GH) effect originating from
the reflection of particles from interfaces \cite{Goos}. Many
works in various graphene-based nanostructures, including single
\cite{Chen15}, double barrier \cite{Song16}, and superlattices
\cite{Chen18}, showed that the GH shifts can be enhanced by the
transmission resonances and controlled by varying the
electrostatic potential and induced gap \cite{Chen15}. Another crucial physical quantity is the group
delay time, which remains among
the important quantities related to the
dynamic aspect of the tunneling process \cite{Hartman, Zhenhua}.
This in fact is often referred to as the Hartman effect, which implies
that for sufficiently large barriers the effective group velocity
of the particle can become superluminal \cite{Olkhovsky, Zhenhua}.

 Moreover, the electronic properties of graphene based
nanostructures can be adjusted by distorting a deformation on the
graphene sample \cite{Haugen,Ni,Mohiuddin,Huang}. Indeed, since
its discovery researchers have conducted extensive research on the
in influence of elastic strain on mechanical and physical properties
of graphene \cite{Sasaki,Maenes}. It is showed that graphene has an
effective young's modulus and simultaneously can reversibly
support elastic strain up to $25\%$ \cite{b5}. Also it is found that  the strain applied
to  graphene allows for producing an energy gap
%provided
%an efficient means to produce an energy
%gap. % as a function of the direction of the strain applied.
%It
%is found that the mechanical strains in graphene
and   changes   the Dirac
points, %. This later  causes Dirac fermions to have
which resulted in having
 asymmetrical effective
Fermi velocities $(v_x^{\eta}, v_y^{\eta})$ for fermions
\cite{Choi,Soodchomshom,Yan1}. Here  $\eta=A,Z$ refers to applied strain  along armchair direction or
zigzag one, respectively.

We address
 the question of how can strain affect the
GH shifts and group delay time in graphene under constraints. %To answer
Then let us
consider a gapped graphene barrier and  apply in the intermediate region a strain along  armchair
and zigzag directions. Solving Dirac equation, we establish
the solutions of energy spectrum for  three regions. From the continuity conditions,  we determine  two transmission probabilities referred to armchair and zigzag directions. These are used to compute
  the corresponding GH shifts and group delay time.
%Subsequently, we numerically study the
%effect of strain along armchair and zigzag directions on the GH
%shifts and group delay time for transmitted Dirac fermions in
%gapped graphene
%under suitable choices of the involved physical parameters.
As results, we show that the strain causes some
changes on the GH shifts and group delay time in transmission
along the armchair direction, but it produces remarkable influence
along the zigzag direction. Consequently, we conclude that
both of these quantities
can be controlled by adjusting the strength of strain along
 each direction.

 The present paper is organized as follows. In section \ref{TTMM}, we formulate our
theoretical problem by writing the corresponding Hamiltonian and
determine the eigenspinors and eigenvalues.
In section \ref{TTPP}, we compute the
transmission probabilities from which we derive the phase shifts.
These are used to obtain
 the GH shifts and group
delay time. We numerically discuss our results by showing different
illustrations under suitable choices of the physical parameters,
in section \ref{NNRR}. Finally, we conclude our results.
% in the final section.

\section{Theoretical model}\label{TTMM}
\begin{figure}[htbp]
    \centerline{\includegraphics[width=0.9\linewidth]
        {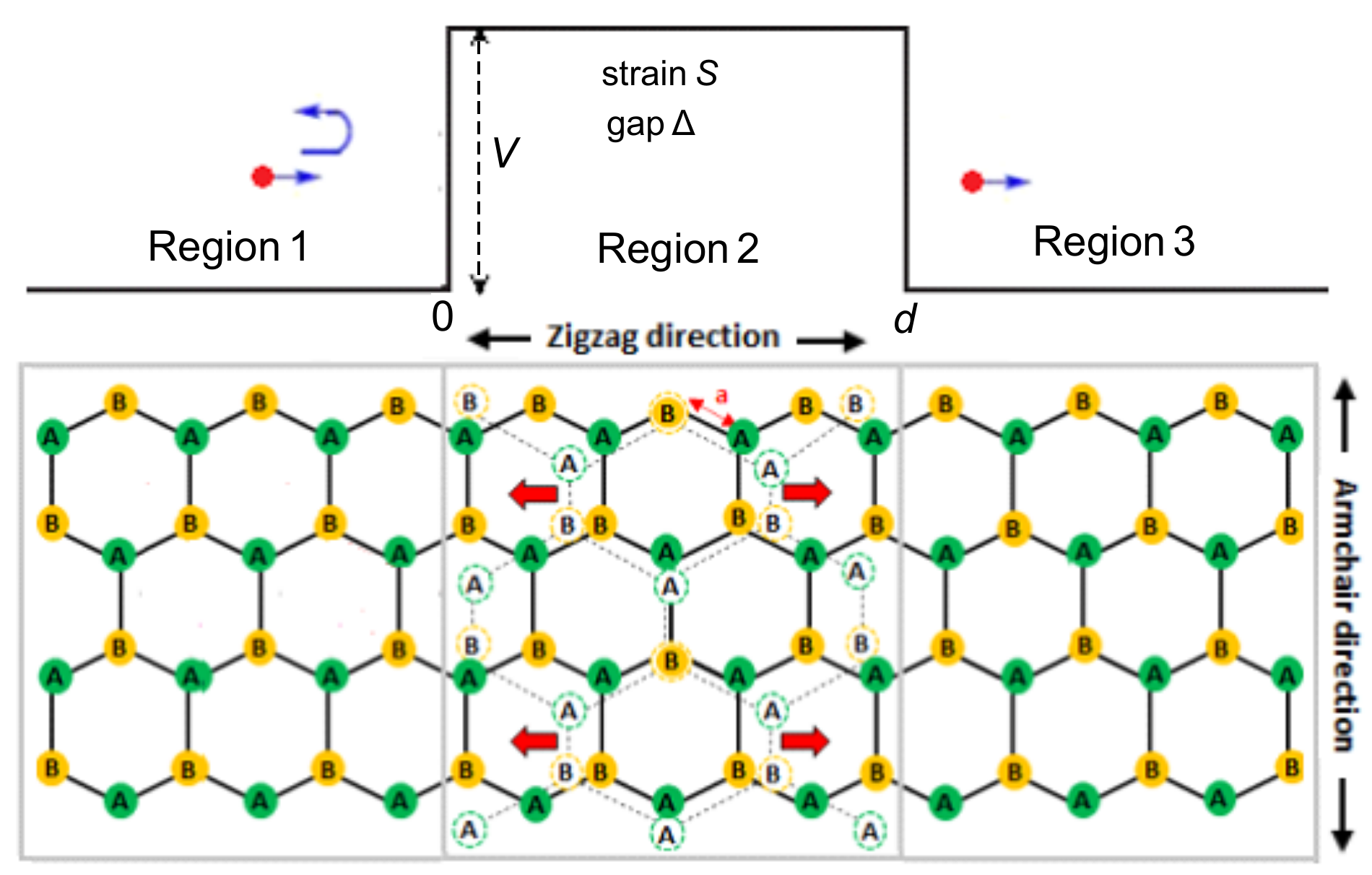}}
    \caption{(color online) Schematically profile of three regions of graphene. The central region is   a gapped graphene subjected to a scalar potential $ V(x) $ together with a  strain strength $ S $ applied armchair ($ y $-axis)
    and     zigzag ($ x $-axis)  directions.}
    \label{fig1}
\end{figure}

We consider a system made of graphene having three regions labeled
by $ j = 1, 2, 3 $ where the intermediate one is subject to a
mass term,  scalar potential and applied strain,
as geometrically presented in FIG.
\ref{fig1}. The mass term $\Delta$ can be
induced by  breaking
%owing to
the
sublattice symmetry through potentials or  spin rotational symmetry
via intrinsic spin orbit coupling \cite{Mele051, Mele052, Sichau19}.
In the framework of the tight-binding approximation, the Hamiltonian governing the motion of the electron in
our system
can be written as
\begin{align}\label{eq1}
H=v_{x}^{\eta}\sigma_{x}p_{x}+v_{y}^{\eta}\sigma_{y}p_{y}
+\left({V}\mathbb{I}_2+{\Delta}\sigma_z\right)\Theta\left(xd-x^{2}\right)
\end{align}
where
$ (\sigma_{x}, \sigma_{y})$ are the usual Pauli matrices, ${\mathbb
I}_{2}$ the $2 \times 2$ unit matrix and $\Theta$ is the Heaviside
step function. The tensional strain affects the Fermi velocity components
to be differently as $v_x^{\eta}$ and
$v_y^{\eta}$ \cite{%Soodchomshom,Yan1,
    Yan2,Wong}.
According
 to the geometry of our system,  we distinguish between applied strain
along  armchair (A) and zigzag (Z) directions. Consequently, we have
the Fermi velocities
\begin{align}\label{eqva2}
 &   v_{x}^{A}= \frac{\sqrt{3}}{2\hbar}a(1-\sigma S)t, \quad v_{y}^{A}= \frac{3}{2\hbar}a(1+S)t'_{3}
\\
&
\label{eqva3}
    v_{x}^{Z}= \frac{\sqrt{3}}{2\hbar}a(1+S)t, \quad
   % \sqrt{4{t'_{1}}^2-{t'_{3}}^2}\\
      v_{y}^{Z}= \dfrac{3}{2\hbar}a(1-\sigma S)t'_{3}
\end{align}
with
$t= \sqrt{4{t'_{1}}^2-{t'_{3}}^2}$,
 $a=0.142$ nm is the distance of the nearest neighbors
without any deformation, $\sigma=0.165$ is the Poisson ratio, and  $S$
is the strain strength. In the tight binding approximation, the only effect
of strain is to modify the altered hopping integral parameter
$t'_{i}$ given by a empirical relation
\begin{align}\label{eqva6}
    t'_{i}= t_{0} e^{-3.37\frac{\delta'_{i}}{a-1}},\qquad  i=1,2,3
\end{align}
resulted from
stretching or shrinking of the distance vectors between the
nearest neighbor carbon atoms \cite{Pereira} and $t_0\approx 2.7$ eV
\cite{Novoselov} is the transfer energy without deformation.
%The
%stress that we applied to the graphene will make a change in the
%nearest neighbor jump parameters, because such a
As a consequence, the strain changes
the distance of nearest neighbors as depicted in FIG. \ref{fig1}
%\ref{fig1} presents the graphene
%atomic
with solid and dashed circles denote sublattices A and
B in undeformed and deformed configurations. As a result
the three nearest neighbor vectors $\delta_{i}$ change to the new ones
$\delta'_{i}$,
such as
\begin{align}
&\label{eqva4}
    \left |\delta'^{A}_{1}\right|= a\left(1-\dfrac{3}{4}\sigma S+\dfrac{1}{4}\ S\right),\quad \left |\delta'^{A}_{3}\right|=a\left(1+S\right)
\\
&
\label{eqva5}
   \left |\delta'^{Z}_{1}\right|=a\left(1+\dfrac{3}{4}\sigma S-\dfrac{1}{4}\  S\right),\quad \left |\delta'^{Z}_{3}\right|=a\left(1-\sigma S\right)
\end{align}
with $ \left |\delta'^{A}_{1}\right|= \left |\delta'^{A}_{2}\right|$
and $ \left |\delta'^{Z}_{1}\right|= \left |\delta'^{Z}_{2}\right| $.

To solve the eigenvalue problem, we proceed by separating variables and
 then write the eigenspinors as $\psi_2(x,
y)=e^{ik_{y}y}\left(\varphi_{2}^{+},\varphi_{2}^{-}\right)^{T}$,
with $k_y$ being  a real parameter that stands for the wave number of the
excitations along the $ y $-axis.
Consequently, the resulting reduced time independent Dirac equation is given by
\begin{align}
\begin{pmatrix}
 V+\Delta- E & -i\hbar(v_x^{\eta}\frac{\partial}{\partial x}+v_y^{\eta} k_{y}) \\
 -i\hbar(v_x^{\eta}\frac{\partial}{\partial x}-v_y^{\eta} k_{y}) & V-\Delta-E \\
\end{pmatrix}
\begin{pmatrix}
  \varphi_{2}^{+} \\
  \varphi_{2}^{-} \\
\end{pmatrix}=0
%E\left(%
%\begin{array}{c}
%  \varphi_{2}^{+} \\
%  \varphi_{2}^{-} \\
%\end{array}%
%\right)
\end{align}
and here the conservation of the momentum $p_y$ has been taken into
account due to the vanishing commutator $[p_y , H]$.
%
%%Now regarding This final the
%%energy eigenvalues $E$ and wavefunction expression $\psi_2(x, y)$
%Now regarding
% region $2$ where the deformation exists,
As a result, in region 2 ($ 0<x<d $) we get the eigenvalues
\begin{equation}\label{eq13}
E=V+s{'}\sqrt{(v_x^\eta \hbar k_x^{\eta})^{2}+(v_y^\eta \hbar
k_{y})^{2}+\Delta^{2}}
\end{equation}
with the sign $s{'}=\mbox{sgn}(E-V)$ refers to conduction and valence bands of region.
The
associated eigenspinors are found to be
\begin{align}
\psi_{2} =\left[a_0
\begin{pmatrix}
\alpha^{\eta}_+ \\
\alpha^{\eta}_- z^{\eta}
\end{pmatrix}
e^{ik_x^{\eta}x}+b_0
\begin{pmatrix}
\alpha^{\eta}_+ \\
 -\frac{\alpha^{\eta}_-}{z^{\eta}}\end{pmatrix}
e^{-ik_x^{\eta} x}\right]e^{ik_{y}y}
\end{align}
where we have set the parameters
$ \alpha^{\eta}_{\pm} $,  $ k^{\eta}_{F} $
and the complex number $ z^{\eta} $
\begin{align}
&\alpha^{\eta}_{\pm}=\left[1\pm\frac{s{'}\Delta}{\sqrt{\Delta^{2}+\hbar^2
\left(v_x^{\eta}k^{\eta}_{F}\right)^{2}}}\right]^{\frac{1}{2}}\\
%&\beta^{\eta}=\left[1-\frac{s{'}\Delta}{\sqrt{\Delta^{2}+\hbar^2
%(v_x^\eta k^{\eta}_{f})^{2}}}\right]^{\frac{1}{2}}
&
k^{\eta}_{F}=\sqrt{\frac{\left(E-V\right)^{2}-\Delta^{2}}
    {(\hbar
v^{\eta}_x)^{2}}}
\\
&
z^{\eta}=
%=s{'}\frac{k_x^{\eta}+i\frac{v_y^{\eta}}{v_x^{\eta}}k_{y}}{\sqrt{(k_x^{\eta})^{2}+(\frac{v_y^\eta}{v_x^\eta}k_{y})^{2}}}=
s{'}
e^{\textbf{-\emph{i}}\phi_{\eta}}, \qquad \phi_{\eta}=\tan^{-1}\frac{v_y^{\eta}k_y}{v_x^{\eta}k_x^{\eta}}
\end{align}
with the wave vector
\begin{equation}\label{eq133}
k_x^{\eta}=s{'}\sqrt{\left(k^{\eta}_{F}\right)^{2}-\left(\frac{v_y^\eta}{v_x^\eta}k_{y}\right)^{2}}
\end{equation}
 $a_0$ and $b_0$ are two constants. The ration $ \frac{v_y^\eta}{v_x^\eta} $
 shows a manifestation of the anisotropy in our system that will play
 a crucial role in the forthcoming analysis.

Regions $1$ and $3$ are assumed to be the infinite pristine
graphene stripes with $S=0$  and an isotropic Fermi velocity, i.e.  $v_{x}=v_{y}=v_{F}$. The eigenspinors
 in region $1$ ($ x<0 $) consists of the incident
and reflected plane waves $ \psi_{1}=\psi_{\text{in}}+ \psi_{\text{re}} $
\begin{equation}\label{555}
\psi_{1}(x,y)=\left[
\begin{pmatrix}
1 \\
 z_0\end{pmatrix}e^{ik_xx}+r
\begin{pmatrix}
1 \\
 -\frac{1}{z_0}\end{pmatrix}e^{-ik_x x
 }\right]e^{ik_{y}y}
\end{equation}
and for region ${\sf 3}$ $(x>d)$, we have $\psi_{3}= \psi_{\text{tr}} $
\begin{equation}\label{777}
\psi_{3}(x,y)=t
\begin{pmatrix}
1 \\
 z_0\end{pmatrix}e^{ik_x x
 }
%\begin{pmatrix}
%1 \\
% -\frac{1}{z_0}\end{pmatrix}e^{-ik_x x}\right]
e^{ik_{y}y}
\end{equation}
%where $\{b_0\}$ is the null vector.
where the incident  wave vector and  $ z_0  $ are given by
\begin{align}
&k_x= \sqrt{k_F^2-k_y^2}\\
& z_0 %=s\frac{k_x +ik_{y}}{\sqrt{k_x^{2} +k_{y}^{2}}}
=s
e^{\textbf{\emph{i}}\phi}, \qquad \phi=\tan^{-1}\frac{k_{y}}{k_x}
\end{align}
with
 $r$ and $t$ denote  the reflection and transmission coefficients, respectively,
$s=\mbox{sgn}(E)$ and
the Fermi wave vector $k_F=\frac{E}{\hbar
v_F}$.

\section{Transport properties}\label{TTPP}

As usual to determine the transmission coefficients  one uses
 the boundary conditions at $x=0$ and $x=d$. This process yields to
 the result
\begin{align}
    t_{\eta}=\frac{e^{ik_x d}\cos\phi_{\eta}\cos\phi}{\cos\phi_{\eta}\cos\phi\cos(k_x^{\eta}d)+i\sin(k_x^{\eta}d)(1+\sin\phi_{\eta}\sin\phi)}
\end{align}
which can be cast to a complex notation
\begin{eqnarray}
t_\eta=\rho e^{i\varphi_{t}^{\eta}}
\end{eqnarray}
 of amplitude $\rho$ and   phase shifts
\begin{align}
\varphi_{t}^{\eta}=\tan^{-1}\left(i\frac{t_\eta^{\ast}-t_\eta}{t_\eta+t_\eta^{\ast}}\right).
\end{align}
%refer to  of the transmission amplitudes.
%Both
%relations will play a crucial rule in computing the GH shift and
%group delay time associated to our system.
%
At this stage  we are ready for
 computing  the corresponding transmission probabilities   $T_\eta$. Indeed, let us
introduce the current  density $J$, which defines
$T_\eta=\frac{J_{\text{tr}}}{J_{\text{in}}}$, with the incident $J_{\text{in}}$ and transmitted $J_{\text{tr}}$  components of $J$. As for our system, we find
\begin{align}
    J=e\upsilon_F \psi^+\sigma_x\psi
\end{align}
giving rise to
 the two transmissions
\begin{align}
    T_\eta=\left|t_\eta\right|^{2}
\end{align}
 and $R_\eta=1-T_\eta$, which resulted from
the conservation law.

 Next, we study the GH shift and
group delay
 by considering some transverse wave vector $k_{y}=k_{y_{0}}$ together with an incident angle  $\phi\left(k_{y_{0}}\right) \in\left[0, \frac{\pi}{2}\right]$, denoted by the subscript 0. An actual finite pulsed electron beam can be
represented as a temporo-spatial wave packet, which is the weighed
superposition of plane wave spinors. Therefore, the wave function
of the incident, refelected at $x=0$ and transmitted wave packets at $x=d$ can be expressed as double
Fourier integral over $\omega$ and $k_y$ \cite{Fatasse2021}
%\cite{Chen3x,Chen4x}
    \begin{align}
    &\label{int1}
    \Phi_{\text{in}}(x,y, t)=\iint f(k_y,\omega)\ \psi_{\text{in}} (x,y) \  e^{-i\omega
        t}\ dk_yd\omega\\
    &\label{int2}
    \Phi_{\text{re}}(x,y,t)=\iint  f(k_y,\omega)\ \psi_{\text{re}} (x,y) \ e^{-i \omega
        t}\ dk_yd\omega
    \\
    &\label{int3}
    \Phi_{\text{tr}}(x,y,t)=\iint  f(k_y,\omega) \ \psi_{\text{tr}} (x,y) \ e^{-i \omega
        t}\ dk_yd\omega
\end{align}
where the three spinors $ \psi_{\text{in}}, \psi_{\text{re}} $ and  $ \psi_{\text{tr}} $ are given in \eqref{555}
and \eqref{777}, respectively.    The frequency of wave is $\omega=E/\hbar$ and
the angular spectral distribution takes the form
%assumed to be  of  Gaussian shape, i.e.
$f(k_y,\omega)=w_ye^{-w_{y}^2(k_y-\omega)^2}$ with   the
half beam width at waist $w_y$ \cite{Beenakker}.
As a result, the total phases
for the reflected and transmitted waves at $x=0, d$
are, respectively,
\begin{align}
    &
\Phi_{r}^{\eta}=\varphi_{r}^{\eta}+k_y-\omega t\\
& \Phi_{t}^{\eta}=\varphi_{t}^{\eta}+k_y-\omega t.
\end{align}
Next,
we use the stationary phase approximation
\cite{Steinberg1, Li1} to establish the  expressions of
 GH shifts and group delay time.
Then the GH shifts in transmissions  are
written as
 \begin{equation}
        S_{t}^{\eta}=- \frac{\partial \varphi_{t}^{\eta}}{\partial
        k_{y}}.
 \end{equation}
 The equation of
motion is obtained using the condition
$\partial\Phi_{t}^{\eta}/\partial\omega=0$ for keeping the good
shape during the propagation, provide the group delay
\begin{align}
       \tau_{t}&= \frac{\partial \varphi_{t}^{\eta}}{\partial
        \omega}+\frac{\partial k_y}{\partial
        \omega}S_{t}^{\eta}\\
% \end{equation}
% we see that there are two contributions to the group delay
% \begin{equation}
    &=
       \tau_{t}^{s}+\tau_{t}^{\varphi}\label{eqq}
 \end{align}
where $\tau_{t}^{\varphi}$ resulted from  time derivative of
phase shifts and  $\tau_{t}^{s}$ is originated from  the
contribution of  $S_{t}$.
% Interestingly, the wave function
%contains two-component spinor so that $\tau_{t}^{\varphi}$ should
%be represented as the average group delay times of two components
%in the following form now on, $\phi$ and $E$ imply the values of
%the central energy and angle of incidence:
As a consequence, we end up with
\begin{equation}
\tau_{t}^{\varphi}=\hbar \frac{\partial \varphi_t^{\eta}}{\partial
E}+\frac{\hbar}{2}\frac{\partial \phi}{\partial E}, \quad
\tau_{t}^{s}=\frac{\sin\phi}{\upsilon_F}S_t^{\eta}.
\end{equation}
These quantities will be numerically computed under suitable conditions of
the physical parameters.
%ThIn fact, we will
%show different plots of the GH shifts and group delay time to
%underline and understand the basic properties of our system.

\section{Results and discussions}\label{NNRR}

The properties of GH shifts $S_t^{\eta}$,
transmission probabilities $T^{\eta}$ and group delay time
$\tau_t^{\eta}/\tau_0$ will be discussed for electrons
transmitting across a barrier in gapped-strained graphene. Here we
introduced the scaled Fermi wavelength $\lambda=\frac{2\pi}{k_F}$
and time scale $ \tau_0 = \frac{d\cos\phi}{v_F}$.
%Before starting, let us fix
% our numerical analysis,
%we first fix some
%requirements needed. Indeed,
Note that according to \eqref{eq133} one should have the condition
\begin{align}\label{eq14}
\left(k_{F}^{\eta}\right)^{2}-\left(\frac{v_y^\eta}{v_x^\eta}k_{y}\right)^{2}\geq 0
\end{align}
in order to have a real wave vector $k_x^{\eta}$ . Then beyond
this, $k_x^{\eta}$ will be imaginary, which physically entails the
evanescence of the wave function inside the barrier. In contrast,
when the strain along armchair and zigzag directions satisfies
\eqref{eq14}, the evanescent wave function exists, but still
propagating inside the transmission region.

{\subsection{The GH shifts in
        transmissions}
\begin{figure}[htbp]
    \centerline{\includegraphics[width=3in,height=2
    in]{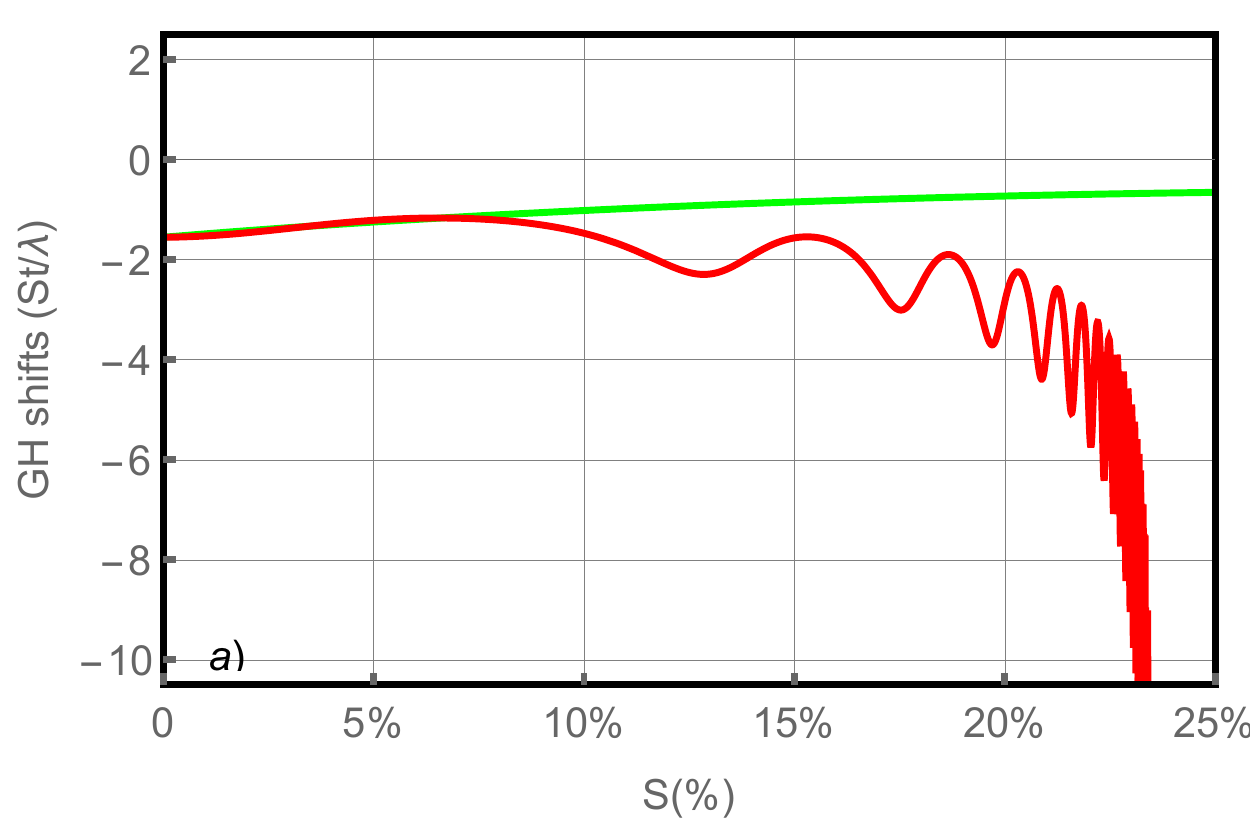}}
\centerline{\includegraphics[width=3in,height=2
    in]{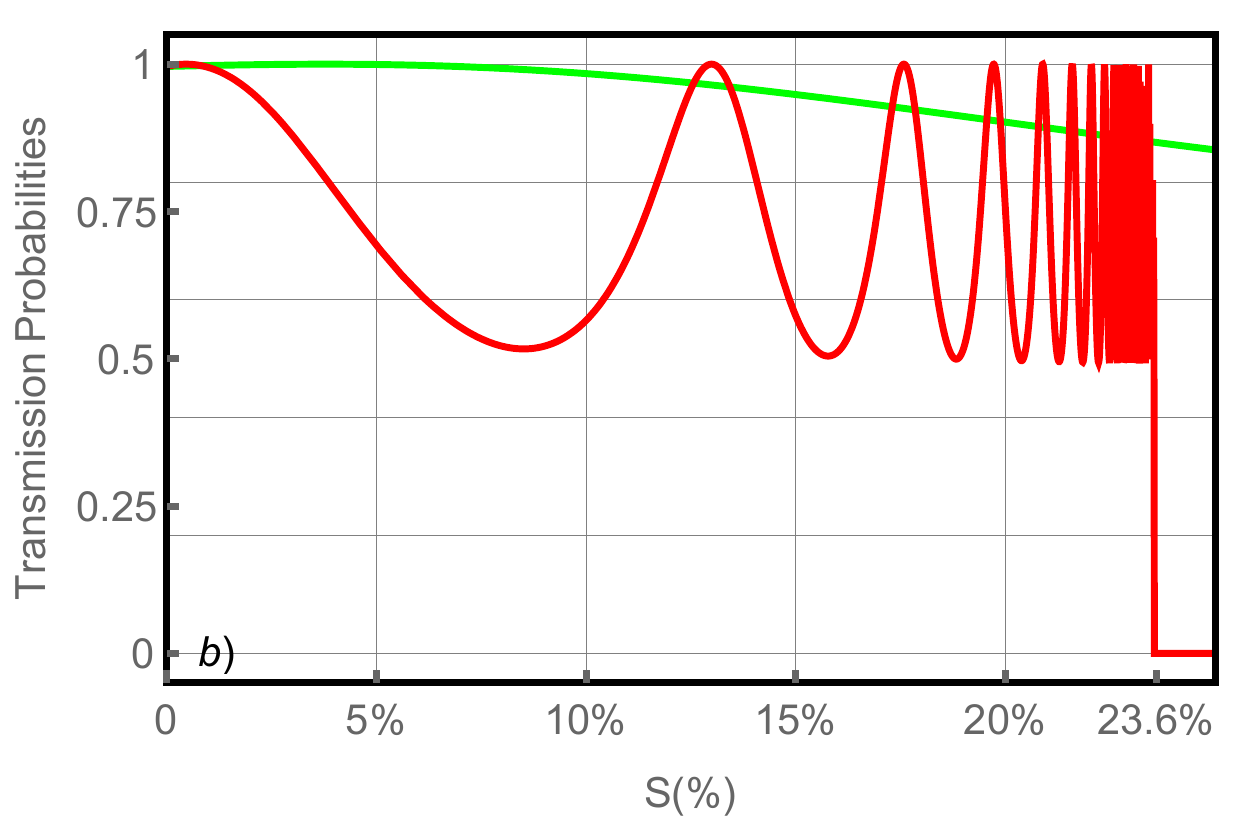}}
    \caption{(color online)   The GH shifts in
transmissions a) and transmission probabilities b) as a function of
the strain strength $S$ for  $V=125$ meV, $E=75$ meV, $\Delta=0$ meV, $d=80$
nm and $\phi=20^{\circ}$. Strain along armchair direction (green line)
and   zigzag (red line).}
        \label{fig2}
\end{figure}
FIG. \ref{fig2} shows the influence of the strain along armchair
and zigzag directions on the GH shifts in transmissions and
transmission probabilities. This has been performed by fixing the
barrier height $V=125$ meV, incident
energy $E=75$ meV, band gap $\Delta=0$ meV, incident angle
$\phi=20^{\circ}$ and barrier width  $d=80$ nm.
From  FIG. \ref{fig2}\textbf{\color{red}{a}}, we observe that  for small values of $S$, the GH shifts in the propagating
mode can be enhanced by transmission resonances. We notice that the GH shifts
decrease by increasing  $S$ for the zigzag case but
increase
in the armchair one.
It is clearly seen  that the GH shifts in transmissions for zigzag  survive beyond the ratio S=23.6\% and vanish at
larger ratio  $S>23.6\%$. In FIG.
\ref{fig2}\textbf{\color{red}{b}}
under the
condition $S>23.6\%$ every incoming state is fully reflected for
the zigzag case.
The
strain along  armchair direction (green line) shows much less
impact on the transmission than strain along  zigzag direction.
This latter
makes the transmissions oscillate
with small amplitudes but high frequencies.

 In FIG. \ref{fig3} we plot the GH shifts and the transmissions
 as a function of the incident energy $E$ in the
strainless $S=0\%$ and strain graphene $S=22\%$ for
 $V_{0}=120$ meV,
$\Delta=0$ meV,  $d$=80 nm and  $\phi$=
$20^{\circ}$.  We observe  that the GH shifts
are closely related to the transmissions. FIG.
\ref{fig3}\textbf{\color{red}{a}}  indicates that the GH shifts
change sign near the Dirac point $E=V$, and become large at
certain resonance points. In fact, the change in sign of the GH
results from the fact that the Dirac point $E=V$ signifies the
transition between the Klein effect $(E<V)$ and the classical
motion $(E>V)$.
% We also notice that the GH shifts in strain
%graphene can have either sign, positive and negative.
The GH shifts present a maximum peak for the zigzag case
compared to armchair case and %whereas the GH shifts
become
constant after certain threshold energy, which is compatible
with a maximum of transmission in FIG.
\ref{fig3}\textbf{\color{red}{b}}. We notice that the oscillating
transmissions decrease for armchair  (green line) and
increase for the zigzag  (red line) compared to
the strainless graphene (blue line).

\begin{figure}[htbp]
    \centerline{\includegraphics[width=3in,height=2
        in]{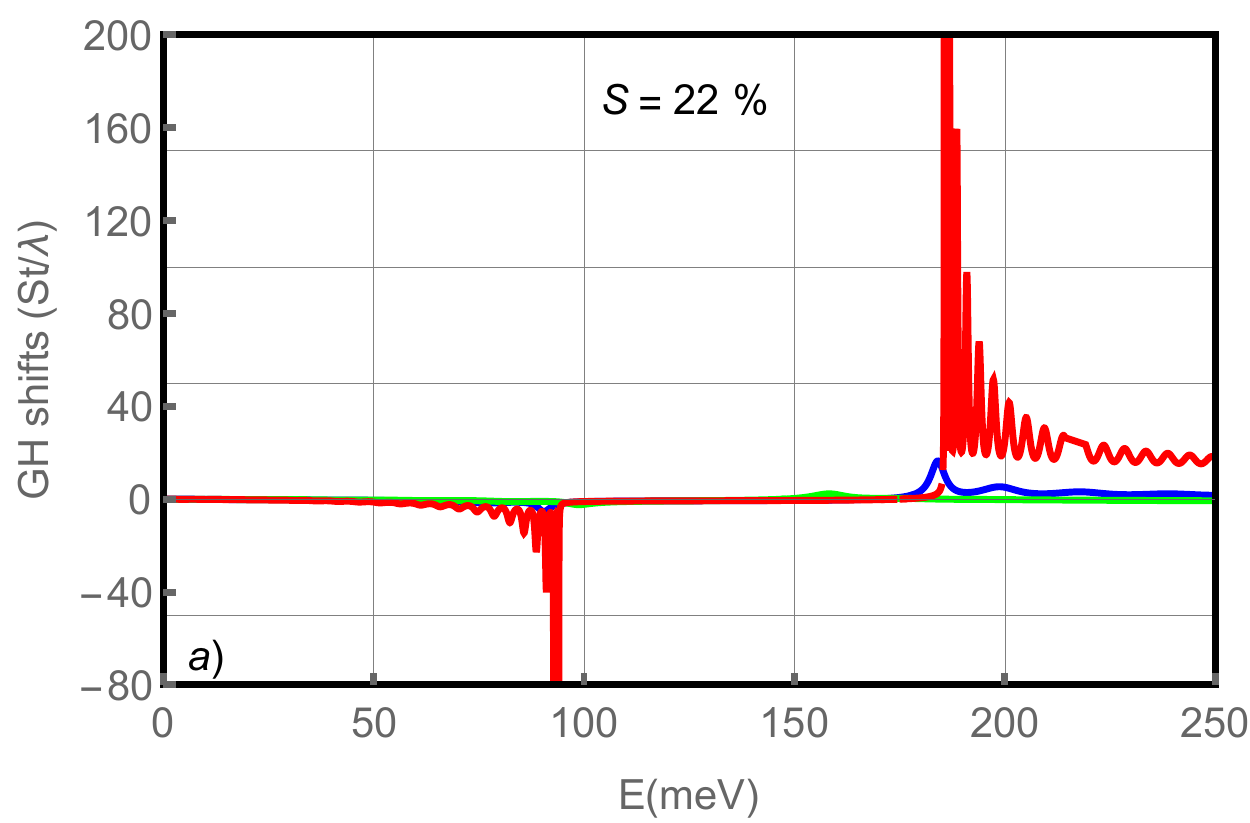}}
    \centerline{\includegraphics[width=3in,height=2
        in]{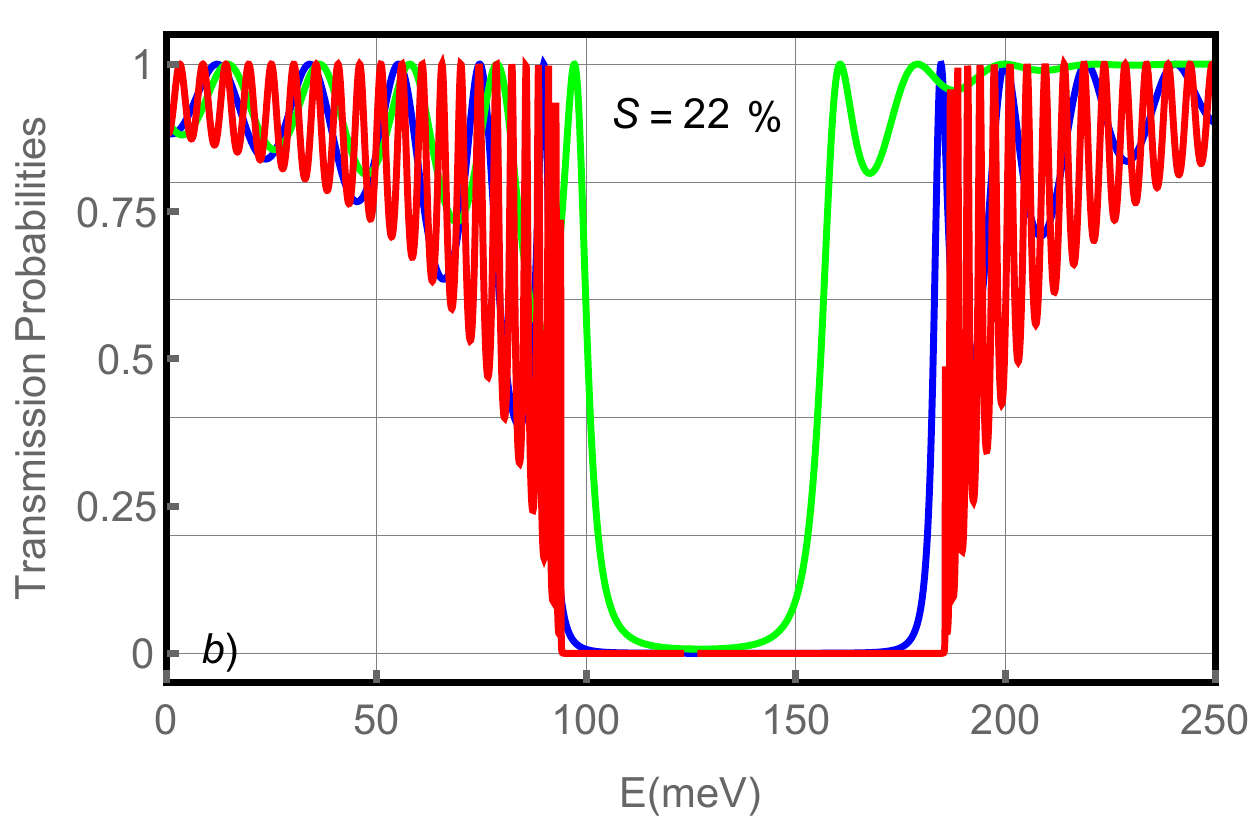}}
    \caption{(color online) The GH shifts in
        transmissions a)  and transmission probabilities  b) as a function of the incident
        energy $E$ for  $V_{0}=120$ meV, $\Delta=0$ meV, $d$=80
        nm, $\phi$= $20^{\circ}$, $S=0\%, 22\%$.
        Strainless (blue line), strain along armchair  direction (green line) and
        zigzag  (red line).}
    \label{fig3}
\end{figure}

\begin{figure}[htbp]
    \centerline{\includegraphics[width=3in,height=2
    in]{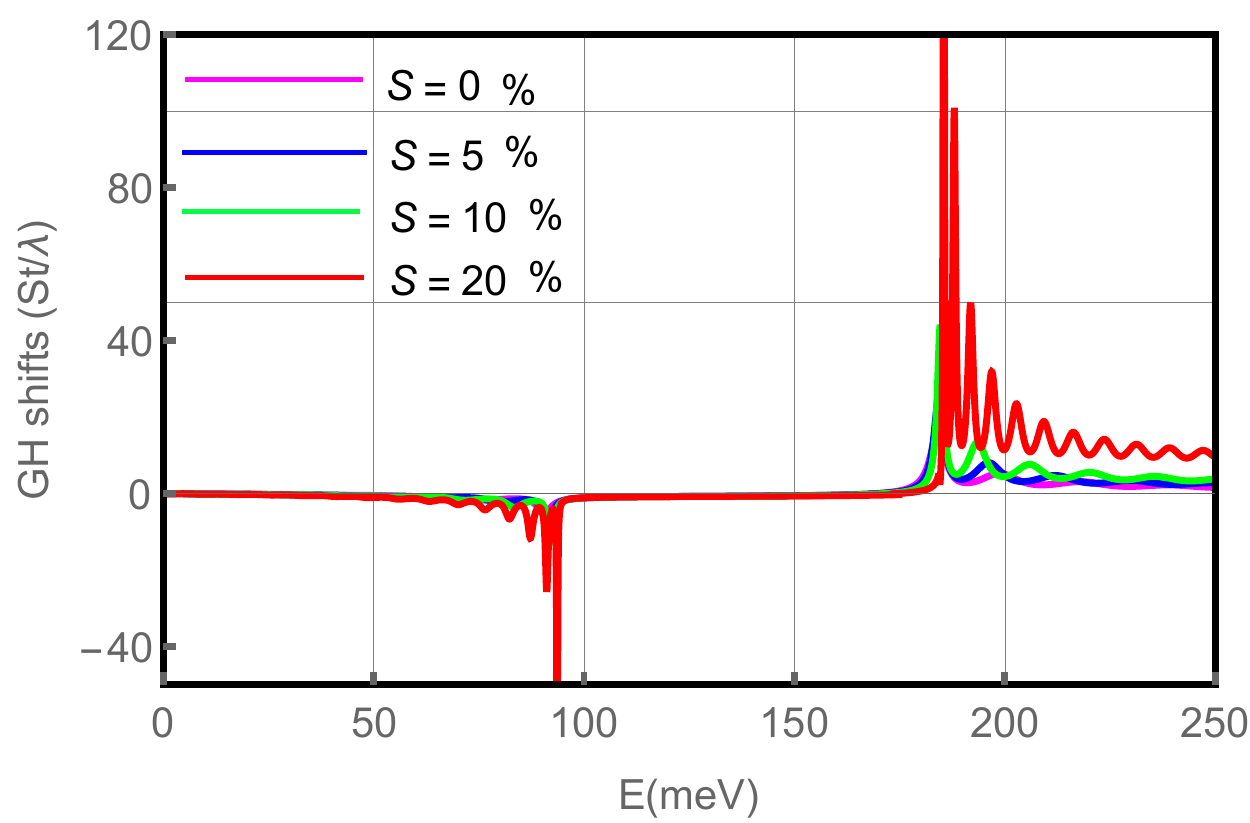}}
    \caption{(color online) The GH shifts in
transmissions for zigzag  direction as a function of the
incident energy $E$ for $V=120$ meV, $d=80$ nm,
$\phi=20^{\circ}$, $\Delta=0$ meV, $S=0\%$ (magenta line), $S=5\%$
(blue line), $S=10\%$ (green line) and $S=20\%$ (red line).}
        \label{fig4}
\end{figure}

 FIG. \ref{fig4} presents the GH shifts in transmissions as a function of the
incident energy
$E$ for strain along zigzag direction with $S=(5\%, 10\%, 20\%)$,
%, which manifests the GH shifts in strained
%graphene  where
$V=120$ meV, $d=80$ nm,
$\phi=20^{\circ}$ and $\Delta=0$ meV. Overall, the GH shifts  evolve in a similar tendency as that in the
strainless case (magenta line) regardless of the strain along  zigzag direction being
$S=5\%$ (blue line), $S=10\%$ (green line) and $S=20\%$ (red
line). However, we observe that
the
GH shifts sensitively depend on the strain strength and show a remarkable
difference between the two values % $S=5\%$,
$S=10\%$ and
$S=20\%$. It turns out that  the strain effect results in the deformation
of the Dirac cones and for that the modulation of GH shifts can be
realized by changing  $S$.

\subsection{Group delay time}

\begin{figure}[htbp]
    \centerline{\includegraphics[width=3in,height=2 in]{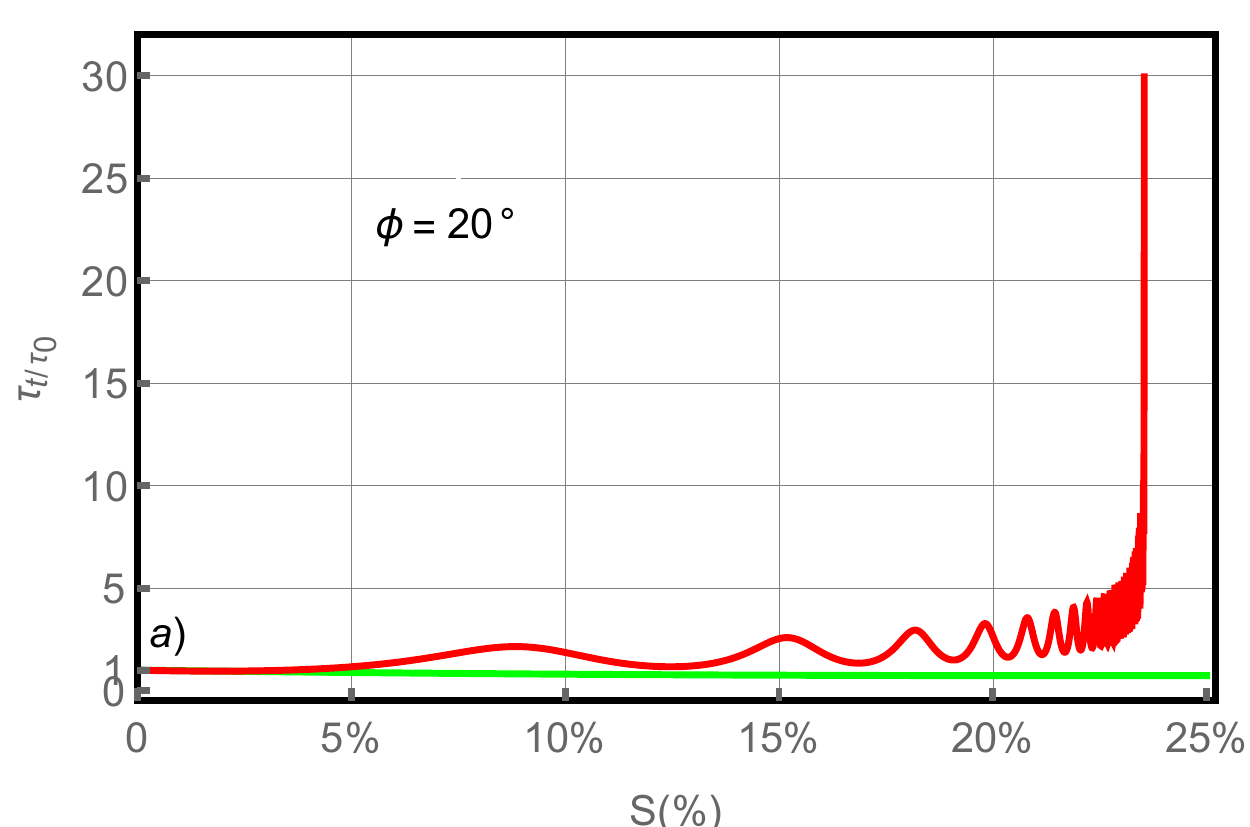}}
    \centerline{\includegraphics[width=3in,height=2 in]{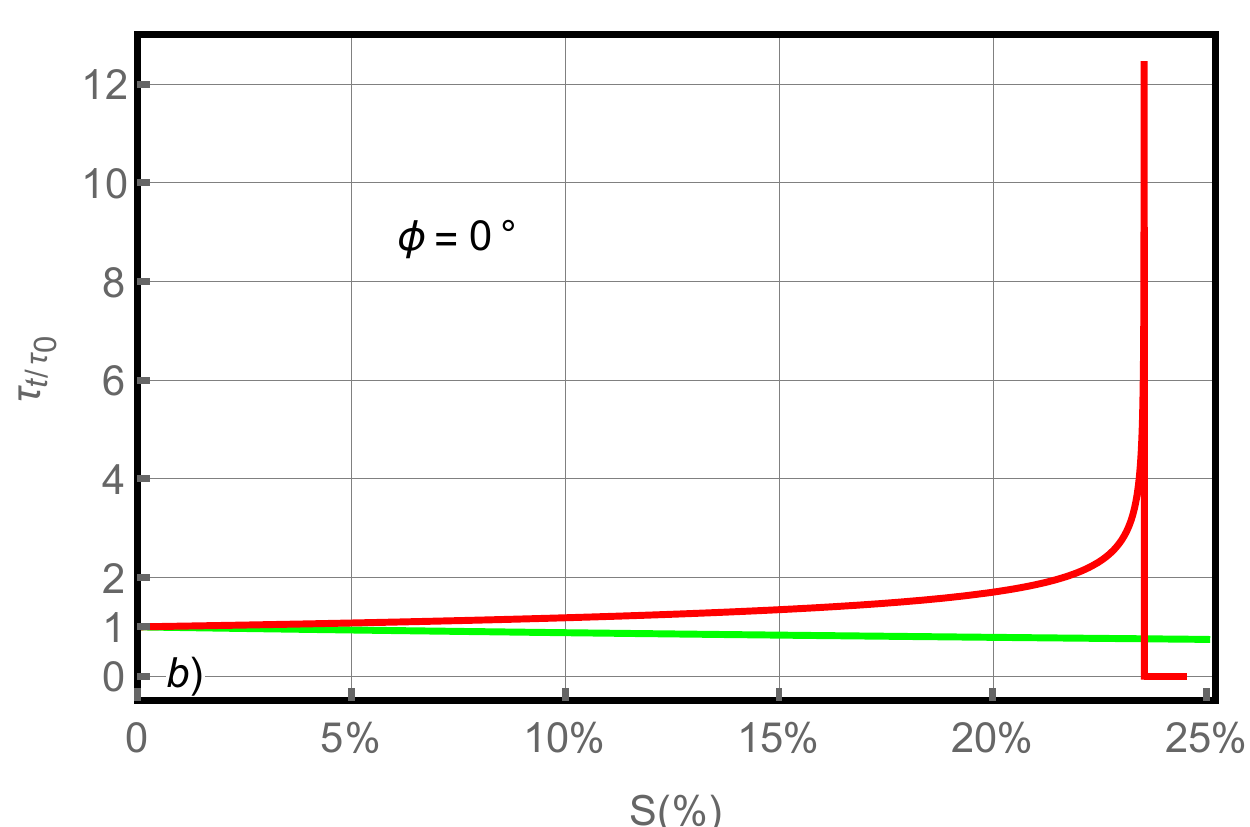}}
    \caption{(color online) The group delay time in transmissions $\tau_t/\tau_{0}$
        as a function of the strain strength $S$ for a):
        $\phi=20^{\circ}$ and b): $\phi=0^{\circ}$, with $E=75$
        meV, $V=123$ meV, $d=100$ nm, $\Delta=0$. Strain along armchair  direction (green line)
        and zigzag  (red line).}
    \label{fig5}
\end{figure}

Now, we investigate the group delay time in transmissions for graphene in the
presence of strain along armchair and zigzag directions. As a
result we will discuss the modulation of group delay by changing
the height of barrier and strain strength $S$
%. The strain effect
%results in the deformation of the Dirac cones, so that the
%modulation of group delay can be realized by changing strain
%modulus $S$ slightly,
 in FIG. \ref{fig5}. For an incident angle $\phi=20^{\circ}$
 in FIG. \ref{fig5}\textbf{\color{red}{a}}, we
observe
%the group delay in the
%propagating mode oscillates for small values of $S$.
%%, and is
%%modulated by the transmission probability T and the GH in the
%%propagating mode can be enhanced by transmission resonances.
%But
the group delay increases by oscillating for strain along zigzag
direction (red line). As  for armchair case (green
line), the group delay is approximately to unity, meaning that the
particles propagate through the barrier with the Fermi velocity
$v_F$ ($\tau_t/\tau_0\simeq 1$).
% {\color{green} decreases with no
%oscillations for armchair case (green line)}. 
FIG.
\ref{fig5}\textbf{\color{red}{b}} shows the group delay  as a
function of strain strength $S$ at normal incidence
$\phi=0^{\circ}$, i.e. $k_y=0$, for both armchair and zigzag
directions in the same choice of  parameters as in FIG.
\ref{fig5}\textbf{\color{red}{a}}.
%$V=123$ meV, $V=75$ meV, $\Delta=0$ meV, $d=100$ nm.
It is clearly seen that
the oscillations of  group delay  disappeared for zigzag
direction but for armchair we still  have the same behavior as in FIG. \ref{fig5}\textbf{\color{red}{a}}.  In addition to these properties, we notice that the absolute
values of the group delay are strongly dependent on the incident
angle.

\begin{figure}[htbp]
    \centerline{\includegraphics[width=3in,height=2 in]{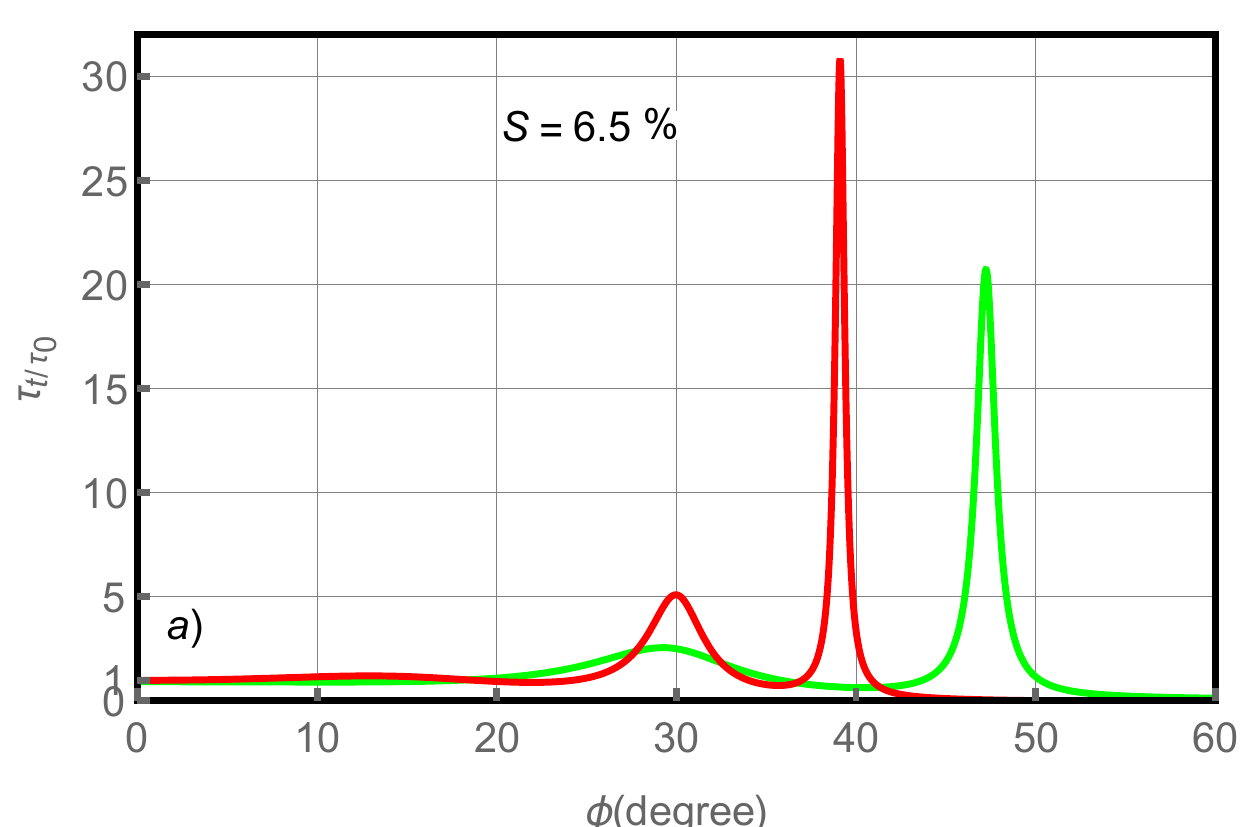}}
    \centerline{\includegraphics[width=3in,height=2 in]{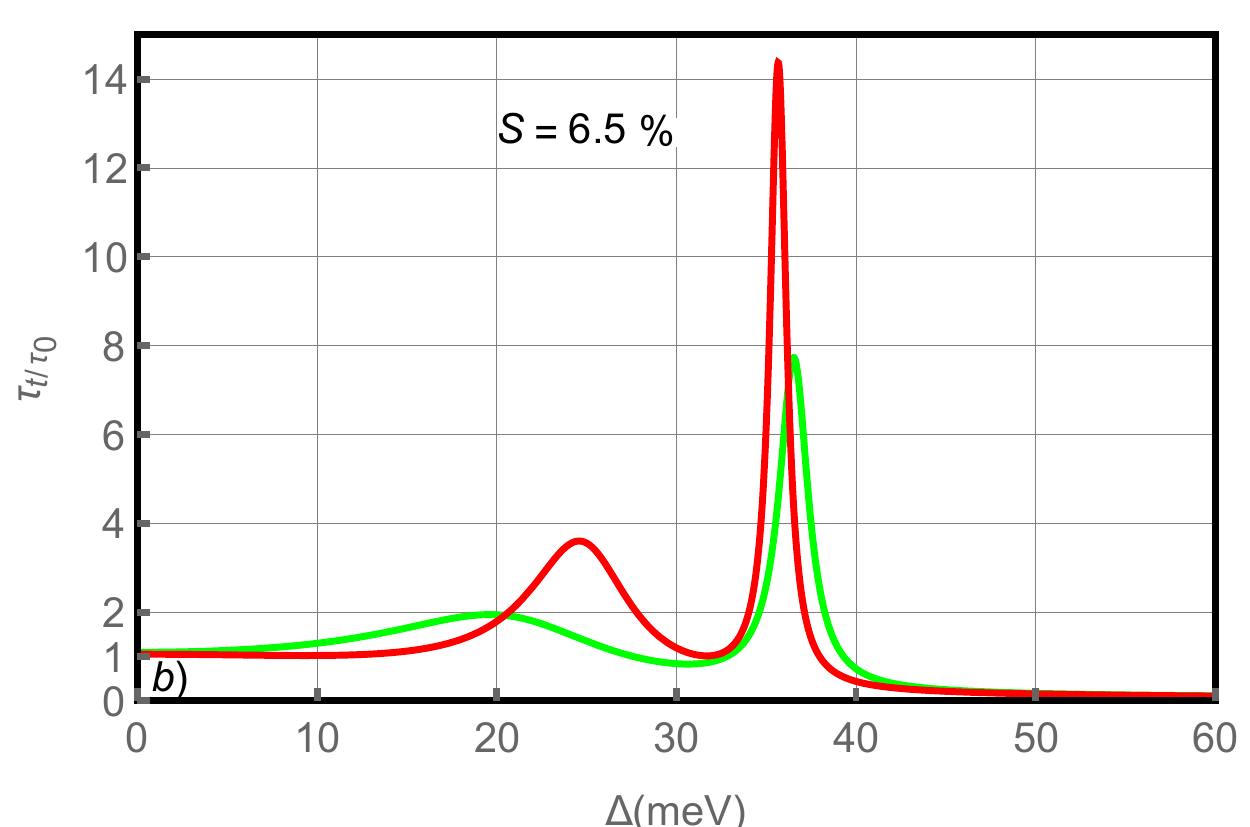}}
    \caption{(color online)  The group delay time in transmissions $\tau_t/\tau_{0}$
        as a function of a)
        the incident angle  $\phi$ for $\Delta=0$
        and b) the gap $ \Delta $ for $\phi=20^{\circ}$. Here we choose
        $V=120$ meV,
        $E=75$ meV $d=100$ nm, $S=6.5\%$. Strain along armchair  direction (green
        line) and zigzag  (red line).}
    \label{fig6}
\end{figure}

 FIG. \ref{fig6}\textbf{\color{red}{a}} we show the
influence of incident angle $\phi$  on the group delay time $\tau_t/\tau_{0}$ in
transmission for strain along armchair and zigzag directions. The
group delay in transmission become mostly constant up to some
value then show sharp picks. It is found that the group delay in
transmission can be enhanced by a certain incident angle. Indeed, by increasing $\phi$, we notice there is
modulation of $\tau_t/\tau_{0}$  for strain along
armchair  and zigzag directions.
%The dependence of $\tau_t/\tau_{0}$ on $\phi$  is shown in FIG.
%\ref{fig6}\textbf{\color{red}{a}},
One sees that $\tau_t/\tau_{0}$ vanishes after
$\phi>45^{\circ}$ for strain along zigzag and $\phi>50^{\circ}$
for  armchair.
%
% Note that graphene is a zero gap semiconductor, despite these great properties suitable, its one
%of the biggest hurdles for graphene to be useful as an electronic
%material \cite{b13}, however, to generate the energy gap is
%crucial for its application in making devices. There are two ways
%to generate the energy gap in monolayer graphene, one requires
%breaking of the translational symmetry \cite{b14}, the other is to
%break the equivalence between the A and B sublattice, which does
%not require any translation symmetry breaking \cite{b15,b16, b17,
%b18}. These fascinating properties of graphene suggest to
%underline the behavior our findings with respect to the induced
%gap. To answer above requirement,
In FIG.
\ref{fig6}\textbf{\color{red}{b}} we show the influence of the
band gap $\Delta$ on the group delay $\tau_{t}/\tau_{0}$ for
strain along armchair (green line) and zigzag (red line) with
$\phi=20^{\circ}$ degree, $V=120$ meV, $E=75$ meV $d=100$ nm,
$S=6.5\%$. We observe that for $\Delta=0$ the particles propagate
through the barrier with the Fermi velocity $v_F $, (i.e.
$\tau_{t}/\tau_{0}=1$). Increasing now $\Delta$, $\tau_{t}/\tau_{0}$ oscillates for the both strain directions.
Additionally, $\tau_{t}/\tau_{0}$ in the case of zigzag strain
exceeds that one of armchair for $\Delta<36$ meV. However,  for
$36\leq\Delta\leq 50$ meV and by increasing   $\Delta$ the former
hierarchy is inverted and therefore $\tau_{t}/\tau_{0}$  for armchair
strain exceeds the zigzag one. Subsequently, as soon as
$\Delta$ increases for $\Delta>50$ meV, $\tau_{t}/\tau_{0}$ will
be frozen, which means that it becomes independent on  $\Delta$.
%It
%has also that increasing the energy gap can lead to more delay
%time which may be beneficial for electronic devices that require
%delay.

\begin{figure}[htbp]
    \centerline{\includegraphics[width=3in,height=2 in]{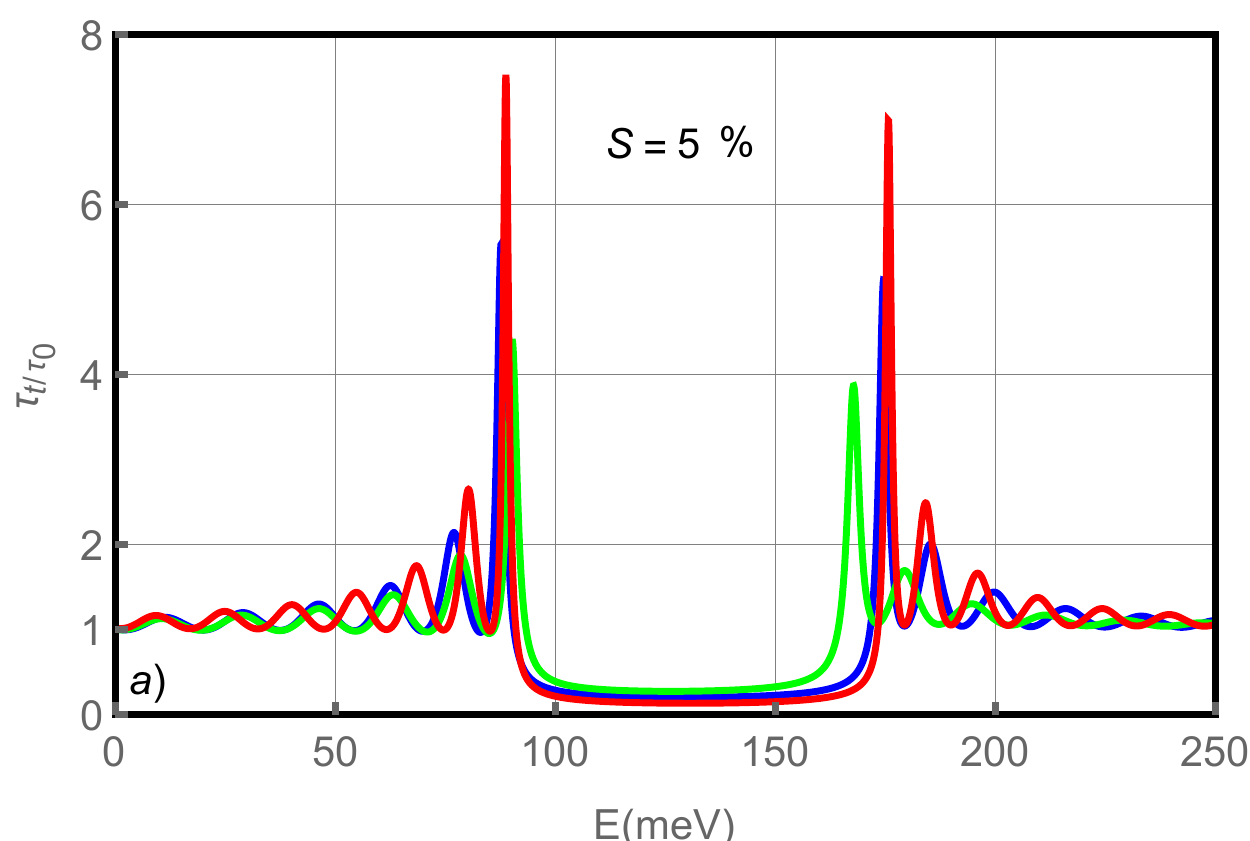}}
    \centerline{\includegraphics[width=3in,height=2 in]{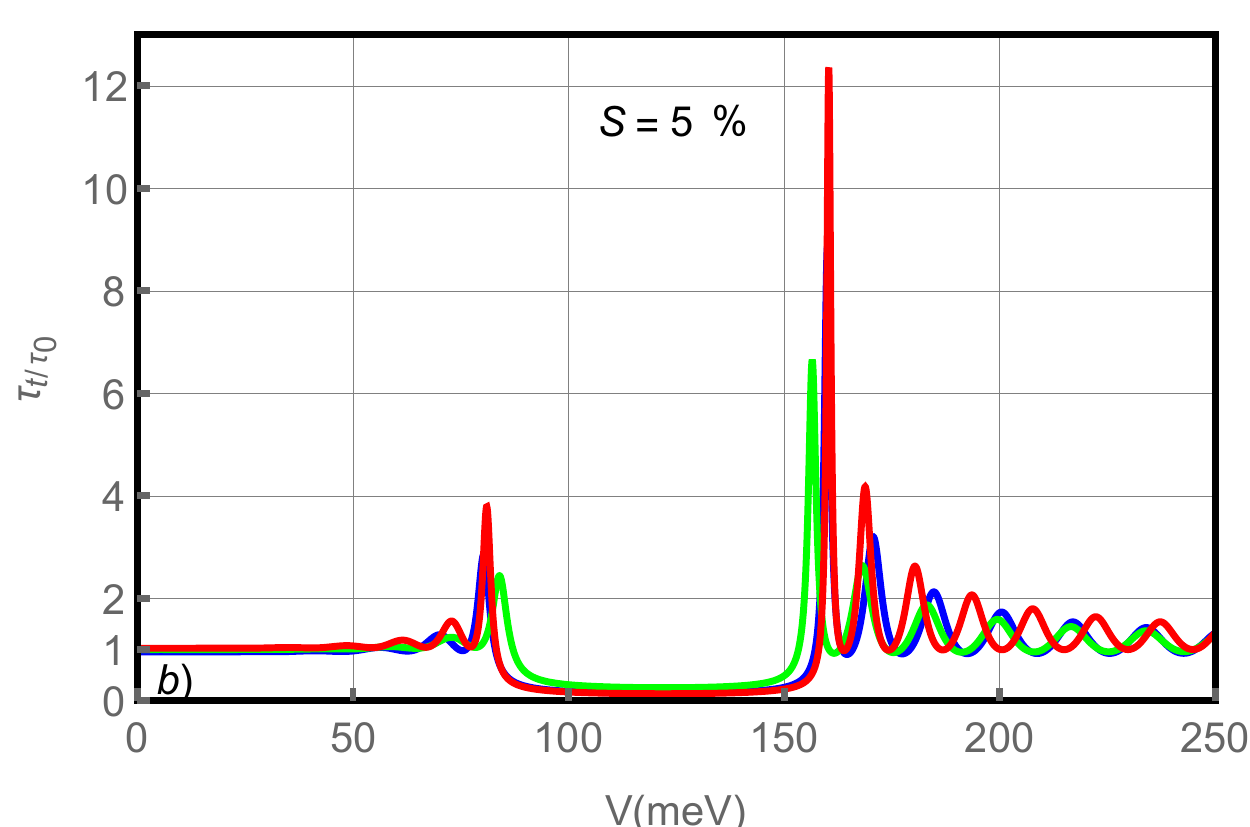}}
    \caption{(color online)  The group delay time in transmissions $\tau_{t}/\tau_{0}$
as a function of
a)
the incident energy $E$ for $V=120$ meV and b) the barrier height $V$
for
$E=120$ meV. Here we choose $d=100$ nm, $\phi=20^{\circ}$, $\Delta=0$, $S=0\%$
(strainless) and $S=5\%$ (strain). Strainless (color blue), strain along
armchair  direction (green line) and zigzag
(red line).}
        \label{fig7}
\end{figure}

 In FIG. \ref{fig7}  we discuss the modulation of group
 delay time in transmissions by varying  the incident energy  and
 the barrier height.
Indeed,
FIG. \ref{fig7}\textbf{\color{red}{a}} presents $\tau_{t}/\tau_{0}$
%the group delay
%time in transmission can be changed by increasing
as a function the incident energy
$E$ for different values of strain strength,   strainless $S=0\%$  and
$S=5\%$ for armchair  and zigzag  directions. It turns out that
the modulation of $\tau_{t}/\tau_{0}$  can be realized
by changing strain $S$. We observe that the amplitude of
oscillations  or peaks increases for zigzag direction and
decreases for armchair  compared to the  strainless case. FIG. \ref{fig7}\textbf{\color{red}{b}} presents $\tau_{t}/\tau_{0}$ as a function of  the
barrier height  $V$. By increasing the strain strength  to $S=5\%$,  $\tau_{t}/\tau_{0}$ decreases for armchair  direction
and becomes less than that for
strainless $S=0$. In contrast,  for  the zigzag direction with $S=5\%$,
$\tau_{t}/\tau_{0}$
increases with respect to $S=0$. The
group delay time in transmission in the propagating mode can be
enhanced by transmission resonances and a null $\tau_{t}/\tau_{0}$ corresponds to a total reflection.

%%%%%%%%%%%%%%%%%%%%%%%%%%%%%%%%%%%%%%%%%%%%%%%%%%%%%%%%%%%%
\section{Conclusion}
%%%%%%%%%%%%%%%%%%%%%%%%%%%%%%%%%%%%%%%%%%%%%%%%%%%%%%%%%%%%%%%
We have studied the  strain effect applied along
armchair and zigzag directions on the GH shifts and group delay
time for transmitted Dirac fermions in gapped graphene through a
single barrier structure. In the first stage,  we have determined the eigenvalues and eigenspinors, which were used to compute the corresponding
transmission probabilities. Subsequently, we have analytically derived
the GH shifts and group delay
time.

 We
have numerically analyzed the GH shifts and group delay time by
considering various choice of the physical parameters. Moreover,
for strain along zigzag direction,  there are increasing of oscillations  in
transmission probabilities, GH shifts and group delay time compared
to the strainless graphene. In contrast, it is found that
such oscillations decrease
for strain along armchair
direction.
%  unlike the zigzag direction the oscillation
We have showed that the group delay time in transmission
approaches unity for a certain critical value of the
barrier height, incident energy, band gap, incident angle and
 barrier width. We have concluded that the group delay time in transmission in the
propagating mode can be enhanced by transmission resonances.

% In this we observe that, the control of GH shift
%and group delay time in transmission by potential, angle incident,
%barrier length, mass term, zigzag directions and armchair
%directions.

%Last but
%not least,
%we should point out that the graphene could provide

May our findings could help to use graphene as a feasible setup to measure the superluminal group delay in solid
state physics.
In addition, tuning the group delay time by scalar potential,
strain strength and gap  could provide some applications in high-speed
graphene-based nanoelectronics \cite{Chen}.

    \end{document}